\documentclass{iaus}
\usepackage{graphicx}
\title[S 262.~~ The Slow Growth of Massive Galaxies] 
{The Slow Growth of Massive Galaxies in Rapidly Growing Dark Matter Halos}

\author[Michael J. I. Brown]   
{Michael J.I. Brown$^1$ and the Bo\"otes Field Collaborations}

\affiliation{$^1$School of Physics, Monash University, Clayton, Victoria 3800, Australia 
\break email: Michael.Brown@sci.monash.edu.au}

\pubyear{2009}
\volume{262}  
\pagerange{1--1}
\jname{Stellar Populations: Planning for the Next Decade}
\editors{G. Bruzual \& S. Charlot, eds.}
\begin{document}

\maketitle

\begin{abstract}
In cold dark matter cosmologies, the most massive dark matter halos are predicted to undergo 
rapid growth at $z<1$. While there is the expectation that massive galaxies will also rapidly
grow via merging, recent observational studies conclude that the stellar masses of the most
massive galaxies grow by just $\sim 30\%$ at $z<1$. We have used the observed
space density and clustering of $z<1$ red galaxies in Bo\"otes to determine how 
these galaxies populate dark matter halos. In the most massive dark matter halos, central
galaxy stellar mass is proportional to halo mass to the power of a $\sim 1/3$ and much
of the stellar mass resides within satellite galaxies. As a consequence, the most
massive galaxies grow slowly even though they reside within rapidly growing dark matter halos.
\end{abstract}

\firstsection 
\section{Introduction}

For plausible cold dark matter cosmologies, the progenitors of the most massive dark matter halos 
form in the early Universe but acquire much of their mass via merging with other halos at $z<1$ 
\cite[(e.g., Lacey \& Cole 1993)]{lac93}. There has thus been an expectation that 
the progenitors of the most massive galaxies will also form when the Universe
is young but acquire much of their mass via galaxy mergers at $z<1$.
This expectation is consistent with the results of most galaxy formation simulations and 
models produced since the 1990s \cite[(e.g., De Lucia et al. 2006)]{del06}.

Recent studies of the $z<1$ galaxy luminosity and stellar mass functions consistently show 
that the most massive galaxies have undergone relatively little growth over the past 7 Gyr 
\cite[(e.g., Bundy et~al. 2006, Brown et~al. 2007)]{bun06,bro07}.
While some of the observations are consistent with zero growth of massive galaxies since $z=1$,
this is not true of all measurements of the galaxy luminosity function \cite[(e.g., Brown et~al. 2007)]{bro07}.
There is also robust evidence that some red galaxies have undergone episodes of star formation 
\cite[(e.g., Trager et~al. 2000)]{tra06}
and galaxy mergers \cite[(e.g., van Dokkum 2005)]{van05} since $z=1$. The most massive red 
galaxies have undergone some growth over the past 7~Gyr, but the exact rate of growth remains a matter of debate.

The contrast between the observed growth of massive galaxies and the predictions of simulations 
has been interpreted as a major problem for the CDM paradigm.
While this may eventually prove to be the case, this is not the case now.
Cosmological galaxy formation simulations and models currently include a number of recipes
for describing the behaviour of baryons, including the cooling and gravitational collapse of
gas, the formation of stars, and the heating of gas by astrophysical sources (e.g., supernovae, quasars).
Given the complexity of modelling the baryons, a disagreement between observations 
and a particular model may have little to do with the validity of the CDM paradigm.
Conversely, agreement between observations and a particular CDM galaxy formation model 
does not necessarily validate the CDM paradigm.

\section{Red Galaxies in Dark Matter}

Why do massive galaxies grow slowly if they reside within rapidly growing dark matter halos?
The evolving luminosity and stellar mass functions of galaxies do not directly address this question.
One can begin to determine why massive galaxies grow slowly, rather than just charting
their evolution, by determining how these galaxies populate dark matter halos.

\begin{figure}
\begin{center}
\includegraphics[width=2.90in]{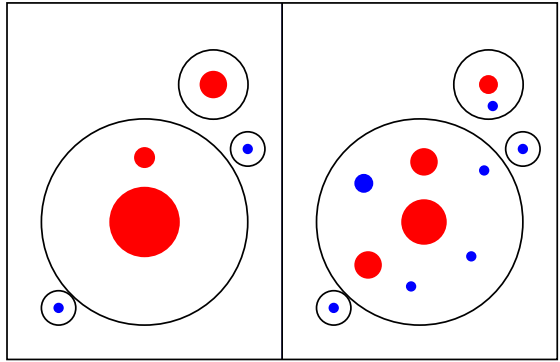}
\caption{A schematic illustrating two galaxy formation scenarios. Both scenarios
have the same halo spatial distribution and mass function, but different scenarios
for how galaxies merge within halos. One can distinguish between these galaxy formation
scenarios by determining how galaxies populate dark matter halos.}
\label{fig:evolscen}
\end{center}
\end{figure}

In Figure~\ref{fig:evolscen} we show schematics of two galaxy formation scenarios with identical 
dark matter distributions. In the first scenario, when dark
matter halos merge the galaxies within these halos merge soon after, resulting in a strong correlation
between galaxy stellar mass and halo mass. In the second scenario, when dark 
matter halos merge the galaxies within these halos orbit the common centre of mass for many billions
of years, and this results in large numbers of satellite galaxies. 
One can distinguish between these galaxy formation scenarios by measuring how galaxies populate dark matter halos.
 
While one can determine dark matter halo masses for individual galaxy groups and clusters, it is not
possible to determine individual halo masses for all galaxies over a broad redshift range. However, 
as one can see in Figure~\ref{fig:evolscen}, the number of close galaxy pairs is a function of the 
number of galaxies per dark matter halo. Also, the spatial clustering of halos is a function of their 
mass, so one can use the large-scale clustering of galaxies to determine the typical halo masses for
various galaxy populations. 

We have measured the luminosity function and clustering of 40,696 $z<1$ red galaxies in the 
Bo\"otes field of the NOAO Deep Wide-Field Survey \cite[(White et~al. 2007, Brown et~al. 2008)], 
and modelled our observations using the halo occupation distribution (HOD) formalism. The HOD
assumes that there is either zero or one central galaxy per dark matter halo, with a population
of satellite galaxies whose spatial distribution (typically) follows the dark matter distribution 
\cite[(e.g., Peacock \& Smith 2000, Zheng 2004)]{pea00,zhe04}.
HOD models can be constrained using the observed space density and clustering of galaxies along with the 
mass function and spatial clustering of dark matter halos from theory. Using the Bo\"otes observations and 
HOD, we have determined how red galaxies populate dark matter halos as a function of galaxy redshift and
luminosity.

\begin{figure}
\begin{center}
\includegraphics[width=3.0in]{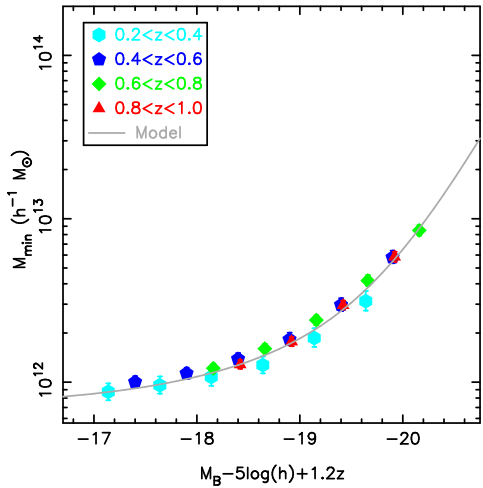}
\caption{The dark matter halo masses of galaxies as a function of absolute magnitude with an 
``evolution correction'' ($1.2z$) added \cite[(Brown et~al. 2008)]{bro08}, so the x-axis 
is effectively a proxy for stellar mass. 
Although galaxy stellar masses and host halo masses evolve, the
relationship between these two quantities shows little evolution at $z<1$.}
\label{fig:dmvsproxy}
\end{center}
\end{figure}

In Figure~\ref{fig:dmvsproxy} we plot the relationship between dark matter halo mass and a proxy  
for central galaxy stellar mass (an evolution corrected luminosity). Although galaxy stellar
masses and dark matter halo masses must evolve between $z=1$ and $z=0$, there is little or no
evolution of the relationship between galaxy stellar mass and host halo mass. In the most
massive dark matter halos, galaxy stellar mass is proportional to halo mass to the power of a
third. As a consequence, if massive dark matter halos double in mass between $z=1$ and $z=0$, 
the stellar masses of the central galaxies within these halos will typically grow by 30\% of the same redshift
range.

\begin{figure}
\begin{center}
\includegraphics[width=2.9in]{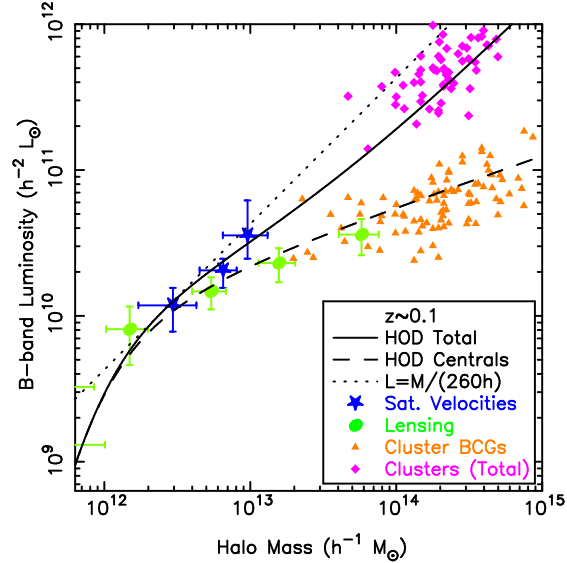}
\caption{The light produced by $z\sim 0.1$ red galaxies as a function of host halo mass \cite[(Brown et~al. 2008)]{bro08}.
In the most massive dark matter halos, central galaxy stellar mass increases slowly with halo mass and the 
bulk of the stellar mass is contained within satellite galaxies rather than a single central galaxy
The data plotted is not from our study but from the work of others \cite[(Lin \& Mohr 2004a, Lin \& Mohr 2004b, 
Mandelbaum et~al. 2006, Conroy et~al. 2007)]{lin04a,lin04b,man06,con07}, who derived halo masses using approaches 
independent of the HOD methodology.}
\label{fig:lightdm}
\end{center}
\end{figure}

In Figure~\ref{fig:lightdm}, we plot the light produced by central and satellite galaxies as a function
of halo mass. Our results are consistent with those derived using independent approaches, including
weak lensing and satellite galaxy velocities. Central galaxy stellar mass increases slowly as a function of 
halo mass in groups and clusters. In the most massive dark matter halos, the bulk of the stellar mass resides
within satellite galaxies and the intra-cluster light, rather than within a single central galaxy. 
This would appear to be consistent with the second scenario illustrated in Figure~\ref{fig:evolscen}, where
galaxy mergers are not an immediate result of dark matter halo mergers. Our results suggest that the 
slow growth of massive galaxies may be consistent CDM cosmologies, but we caution that this should not
confused with a robust test of the $\Lambda$CDM paradigm.

While we now have a description of how massive galaxies are growing within dark matter halos at $z<1$, 
 key questions remain unanswered. Several proposed mechanisms for the regulation of star formation
have a strong dependence on halo mass (e.g., virial shock heating), so important insights could be gained by  
determining how blue galaxies populate dark matter halos. 
While it is often assumed that the bulk of red galaxy stellar mass growth is via galaxy mergers, the most 
massive red galaxies grow so slowly at $z<1$ that star formation may make a significant contribution.
Only the last 30\% of massive galaxy growth has been accurately measured, and the bulk of massive
galaxy growth must take place between $z\sim 1$ and the peak of star formation at $z\sim 3$. 
Finally, most halo masses for $z>0.5$ galaxies depend upon models of 
the clustering and mass function of dark matter halos, and observational confirmation of these
masses is required.

\section{Summary}

In CDM cosmologies the most massive dark matter halos undergo rapid growth via merging between $z=1$ and $z=0$.
At first glance, the slow growth of massive galaxies at $z<1$ appears contrary to the CDM paradigm, but this 
is not necessarily the case. 
Using the observed space density and clustering of red galaxies, we have been 
able to model how red galaxies populate dark matter halos. In the most massive dark matter halos, central
galaxy stellar mass is a weak function of host halo mass and the much of the stellar mass resides within satellite
galaxies rather than a single central galaxy. As a consequence, massive galaxies grow slowly even though they
reside within rapidly growing dark matter halos.

\end{document}